\documentclass[aps,showpacs,prc,twocolumn]{revtex4}
\usepackage{graphicx}
\usepackage{amssymb}
\begin{document}

\title{Absence of decoherence in the complex potential approach to nuclear scattering}

\author{Alexis Diaz-Torres}
\affiliation{Department of Physics, Faculty of Engineering and
Physical Sciences, University of Surrey, Guildford, Surrey GU2 7XH,
United Kingdom}

\begin{abstract}
Time-dependent density-matrix propagation is used to demonstrate, in a schematic model of an open quantum system, that the complex potential approach and the Lindblad dissipative dynamics are \emph{not} equivalent. While the former preserves coherence, it is destroyed in the Lindblad dissipative dynamics. Quantum decoherence is the key aspect that makes the difference between the two approaches, indicating that the complex potential model is inadequate for a consistent description of open quantum-system dynamics. It is suggested that quantum decoherence should always be explicitly included when modelling low-energy nuclear collision dynamics within a truncated model space of reaction channels.   
\end{abstract}
\pacs{03.65.Yz, 03.65.Xp, 03.65.Nk, 24.10.-i}
\maketitle
 
Quantum coherence manifests itself through quantum interference effects in quantum dynamical systems, which is clearly demonstrated in double-slit experiments with single electrons. This property of matter is diminished (decoherence) when an external environment interacts with the system \cite{Lucia,Sonnentag,Zimmermann}. Quantifying the role and importance of decoherence in quantum systems is now pervasive in various branches of physics and chemistry, including studies of quantum measurement and quantum information \cite{Zurek}. The concept of a {\em reduced} (but not closed) quantum system evolving in the presence of couplings to an environment of complex states is common throughout disciplines.

The schematic model refers to the scattering of a wave-packet $\psi_{k_0}(x)$ (with the mean wave-number $k_0$) off a potential barrier $V(x)$. This is considered a reduced quantum system, which irreversibly interacts with a complex environment of excluded degrees of freedom. Of interest is the effect of such an environment on the reduced quantum-system dynamics. It is here investigated using two descriptions: (i) the complex potential approach \cite{Wigner,Feshbach,Volya}, and (ii) the Lindblad dissipative dynamics \cite{Lindblad,Sandulescu}. The key question addressed in this paper is: Are these approaches equivalent? A common view in the nuclear physics community is that they are. This general belief inhibits research into quantum decoherence which is at the same time highly topical in other areas of physics and chemistry. My model calculations clearly demonstrate that quantum decoherence makes the difference between the two descriptions, and is \emph{not} accounted for in the complex potential model. The methodology will be highlighted first. Afterwards, the calculations are presented and discussed, and finally a summary is given. 

The Liouville-von Neumann master equation $d \hat{\rho}/d t = \mathcal{L}\, \hat{\rho}$ dictates the time evolution of the reduced-system density-matrix operator $\hat{\rho}(t)$. Its initial value describes a pure state, and is determined by the initial wave-packet as $\hat{\rho}(0)=|\psi \rangle \langle \psi |$. The Liouvillian $\mathcal{L}\, \hat{\rho}$ is different in the two descriptions studied:

\begin{description}
\item \textnormal{(i)} In the complex potential approach \cite{Wigner}, this is
\begin{equation} 
\mathcal{L}\, \hat{\rho}\, = \, -\frac{i}{\hbar}( \, \hat{H}_{eff} \hat{\rho} \, - 
\, \hat{\rho} \hat{H}_{eff}^{\dag} \, ),
\label{eq1}
\end{equation}
where the effective non-Hermitian Hamiltonian $\hat{H}_{eff} = \hat{H}_{s} - i\, W(x)$, being $\hat{H}_{s} = \hat{T} + V(x)$ the Hermitian Hamiltonian of the reduced system ($\hat{T}$ is the kinetic energy operator) and $W(x) > 0$ describes the irreversible environmental interaction(s).     
\item \textnormal{(ii)} Irreversibility in dynamics of an open quantum system can be consistently described by the Lindblad master equation \cite{Lindblad}. Here the Liouvillian reads as
\begin{eqnarray} 
\mathcal{L}\, \hat{\rho} &=& -\frac{i}{\hbar}\, [\, \hat{H}_s, \hat{\rho}\,] \, + \, 
\nonumber \\ 
&+& \sum_\alpha \bigl( \hat{\mathcal
C}_\alpha \, \hat{\rho} \, \hat{\mathcal C}_\alpha^{\dag} -
\frac{1}{2} \bigl[\hat{\mathcal C}_\alpha^{\dag} \, \hat{\mathcal
C}_\alpha ,\hat{\rho} \bigl]_{+} \bigl)\, 
\label{eq2}
\end{eqnarray}
where $[\ldots]$ and $[\ldots]_{+}$ denote the commutator and anti-commutator, respectively. Each $\hat{\mathcal C}_\alpha$ is a Lindblad operator for a 
dissipative coupling, physically motivated according to the specific problem. 
We use a \emph{spontaneous emission} Lindblad operator \cite{Irene}, \emph{i.e.},   
$\hat{\mathcal C}_{21}=\sqrt{\gamma_{xx}^{21}} |2 \rangle \langle 1|$ that describes a decay from the reduced-system state $|1 \rangle$ to the environmental state $|2 \rangle$. (These states are assumed to be orthonormal.) State $|2 \rangle$ is not a reaction channel, but an auxiliary state \cite{Irene} supplying a probability drain only. This state mocks up a high density of complex states (environment) and irreversibly absorbs probability from the reduced system. The absorption rate to state $|2 \rangle$ is given by $\gamma_{xx}^{21} = W(x)/\hbar$ 
where $W(x)$ is taken as in approach (i). Thus, approaches (i) and (ii) refer to a single reaction channel problem.  
\end{description}

The Liouville-von Neumann master equation with either (\ref{eq1}) or (\ref{eq2}) is then represented in a grid-basis \cite{Kosloff2}. The density-matrix elements obey the following equations:

\begin{description}
\item \textnormal{(a)} In the approach (i),
\begin{equation}
\dot{\rho}_{x x'} \, = \, -\frac{i}{\hbar}\, [\, \hat{H}_s, \hat{\rho}\,]_{x x'} \, + \,
(\mathcal{L}_D\, \hat{\rho})_{x x'},  
\label{eq3}
\end{equation}       
where 
\begin{equation}
(\mathcal{L}_D\, \hat{\rho})_{x x'} \, = \, -\frac{1}{\hbar}\, ( \, W(x) + W(x') \,)\, 
\rho_{x x'},
\label{eq4}
\end{equation}
\item \textnormal{(b)} Whereas in the approach (ii),
\begin{equation}
\dot{\rho}_{x x'}^{11} \, = \, -\frac{i}{\hbar}\, [\, \hat{H}_s, \hat{\rho}\,]_{x x'}^{11} \, + \,
(\mathcal{L}_D\, \hat{\rho})_{x x'}^{11},  
\label{eq5}
\end{equation}
\begin{equation}
\dot{\rho}_{x x'}^{22} \, = \, (\mathcal{L}_D\, \hat{\rho})_{x x'}^{22},  
\label{eq6}
\end{equation}
and the elements $\rho_{x x'}^{12}$ and $\rho_{x x'}^{21}$ obey equations of motion like (\ref{eq6}). The dissipative terms are 
\begin{eqnarray}
(\mathcal{L}_D\, \hat{\rho})_{x x'}^{k l} &=&
\delta_{k l}\, \sum_{j=1}^{2} \sqrt{\gamma_{xx}^{kj} \, \gamma_{x'x'}^{kj}} \,  
\rho_{x x'}^{j j} \, - \, \nonumber \\ 
&-&\frac{1}{2}\, \sum_{j=1}^{2}\, ( \, \gamma_{xx}^{jk} \, + \, \gamma_{x'x'}^{jl} \,)\, 
\rho_{x x'}^{k l},
\label{eq7}
\end{eqnarray}
where $(k,l=1,2)$ and $\gamma_{xx}^{kk} = \sum_{j \ne k} \gamma_{xx}^{jk}$ \cite{Saalfrank}. It makes the second term of (\ref{eq7}) equal to (\ref{eq4}) for the elements $\rho_{x x'}^{11}$ that describe the reduced system. Thus, the first term of (\ref{eq7}) makes the difference between the two approaches, which is essential as shown below. It is worth mentioning that the elements $\rho_{x x'}^{22}$ absorb probabilities only, whilst $\rho_{x x'}^{12}$ and $\rho_{x x'}^{21}$ are always zero. These are initially zero like $\rho_{x x'}^{22}$. 
\end{description}

Having obtained the dynamical evolution of the density matrix, expectation values of an observable $\hat{\mathcal O}$ result from the trace relation $\langle
\hat{\mathcal O (t)} \rangle  = \textnormal{Tr}(\hat{\mathcal O}\hat{\rho}(t))$.

\begin{figure}
\begin{center}
\includegraphics[width=0.40\textwidth,angle=0]{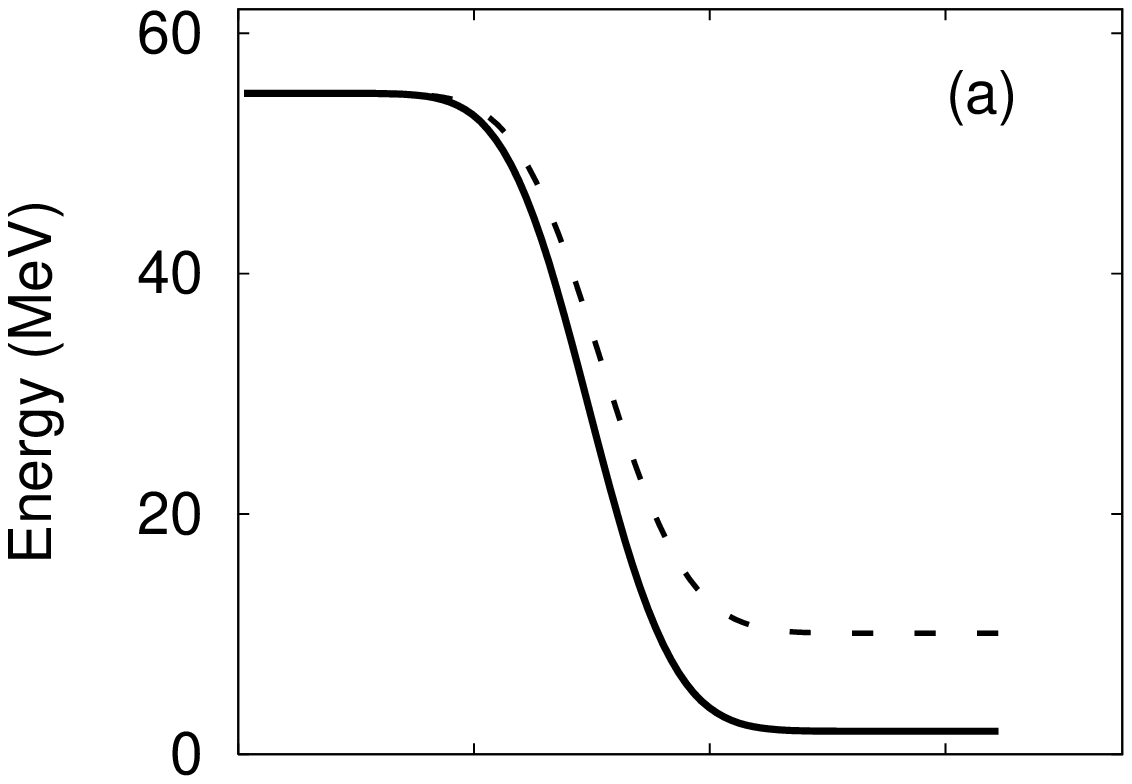} \\ 
\includegraphics[width=0.40\textwidth,angle=0]{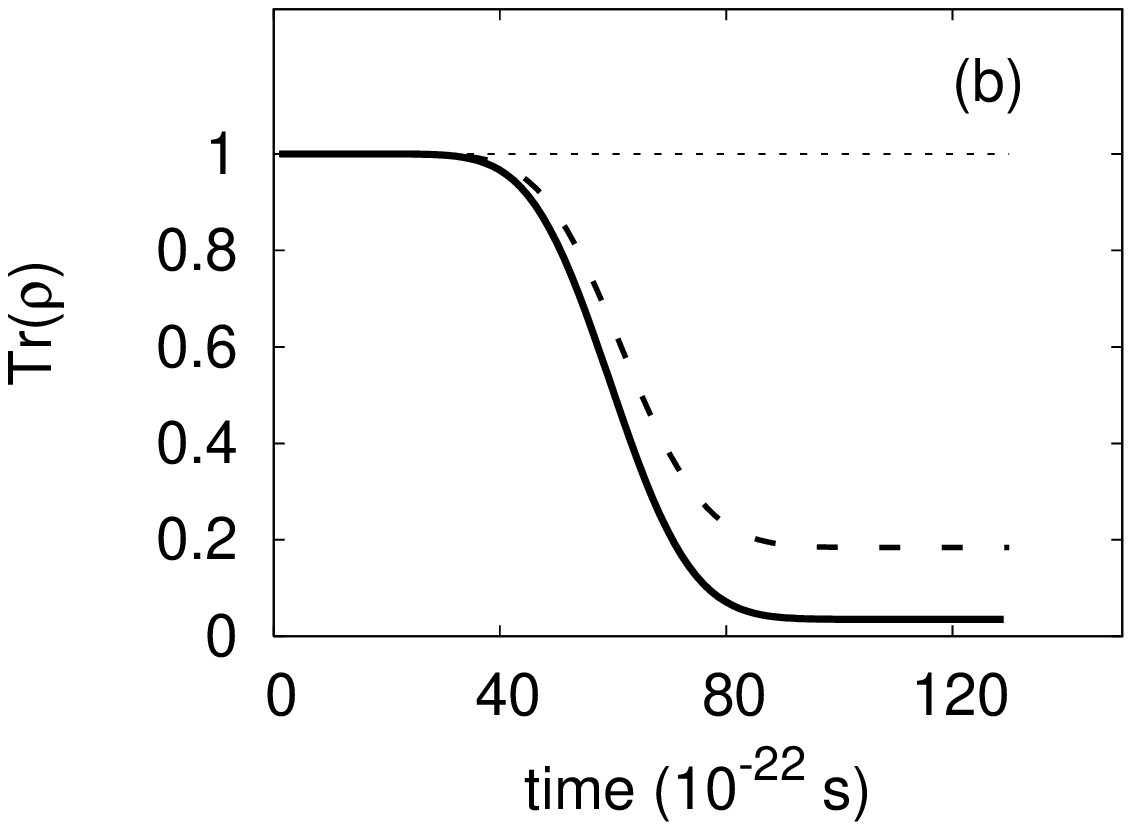}
\caption{(a) The energy mean value of the reduced system as a function of time in the complex potential approach (solid line) and in the Lindblad dissipative dynamics (dashed line). (b) The same but for the trace of the density matrix. Lindblad's dynamics preserves the total probability (dotted line), \emph{i.e.}, the sum of the probability in the reduced system and in the environment.} 
\label{Figure1}
\end{center}
\end{figure}

\emph{Results and discussion.} In the model calculations the grid (-50 to 50 fm) was evenly spaced with 256 grid-points. The potentials $V(x)$ and $W(x)$ are the Eckart potential \cite{Eckart} and a Gaussian function, both centered at zero, 
\emph{i.e.}, $V(x)=V_0 \, cosh^{-2}(x/a_{0})$ and $W(x)=W_0 \, exp[-x^2 / 2 \sigma_{0w}^2]$. The parameters are ($V_0,a_0)\equiv$ (61.1 MeV, 1.2 fm) and ($W_0,\sigma_{0w})\equiv$ (5 MeV, 1.5 fm). A minimal-uncertainty Gaussian wave-packet describes the relative motion of two objects with mass numbers $A_1=16$ and $A_2=144$. The wave-packet is initially centered at $x_0 = 25$ fm, with width $\sigma_0 = 5$ fm, and was boosted towards the potentials with the appropriate average kinetic energy for the initial total energy $E_0=55$ MeV.
 
The density-matrix was propagated in time using
the Faber polynomial expansion of the time-evolution operator \cite{Faber}, and the Fourier method \cite{Kosloff2} for the commutator between the kinetic
energy and density operator. The time step for the density-matrix propagation was $\Delta t = 10^{-22}$ s. The numerical accuracy of the propagation was checked using
a time-dependent calculation without the dissipative term $\mathcal{L}_D\, \hat{\rho}$ in (\ref{eq3}) or (\ref{eq5}). It was confirmed that the
trace and purity of the density-matrix, Tr$(\hat{\rho})$ =
Tr$(\hat{\rho}^2)=1$, and the expectation value of the system energy
Tr$(\hat{H} \hat{\rho})$ were maintained with high accuracy over the
required number of time steps, typically 130 when the centroid of the 
recoiled body of the wave-packet reaches 25 fm. 

\begin{figure}
\begin{center}
\includegraphics[width=0.40\textwidth,angle=0]{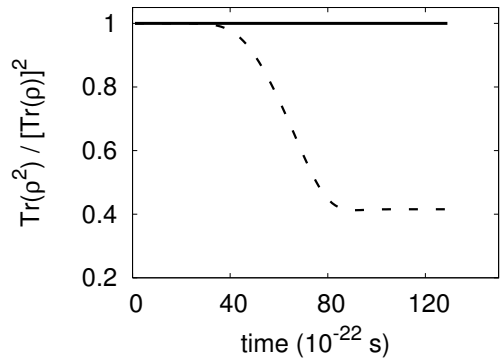}
\caption{The same as in Fig. \ref{Figure1}, but for the measure of coherence \cite{Probe} in 
the reduced system. While the complex potential approach preserves coherence (solid line), it is destroyed in the Lindblad dissipative dynamics (dashed line).} 
\label{Figure2}
\end{center}
\end{figure}

Figure \ref{Figure1} shows the time evolution of the reduced-system energy (Panel a) and the trace of the density-matrix (Panel b) using both the complex potential approach (solid line) and the Lindblad dynamics (dashed line). These quantities show similar features, their values declining in time as expected. The total probability is mantained in Lindblad's dynamics (dotted line), whilst the complex potential approach \emph{cannot} guarantee its preservation. The disagreement of the solid and dashed lines is due to the first term of (\ref{eq7}). Removing it, the two approaches provide the same results. However, this term is crucial, making the two approaches physically different due to the quantum decoherence, as shown in Fig. \ref{Figure2}.   

The measure of coherence in the reduced system is the ratio Tr$(\hat{\rho}^2)/[$Tr$(\hat{\rho})]^2$ \cite{Mueller,Probe}. Its time evolution is presented in Fig. \ref{Figure2} for the complex potential approach (solid line) and the Lindblad dynamics (dashed line). The former preserves coherence, whilst the Lindblad dynamics results in decoherence. The complex potential leads to a \emph{coherent} damping of all the elements of the density matrix, affecting only the amplitude of the off-diagonal elements. However, the first term of (\ref{eq7}) results in \emph{dephasing}, destroying the initial phase relationship between the off-diagonal elements. This \emph{is} quantum decoherence. 

\begin{figure}
\begin{center}
\includegraphics[width=0.40\textwidth,angle=0]{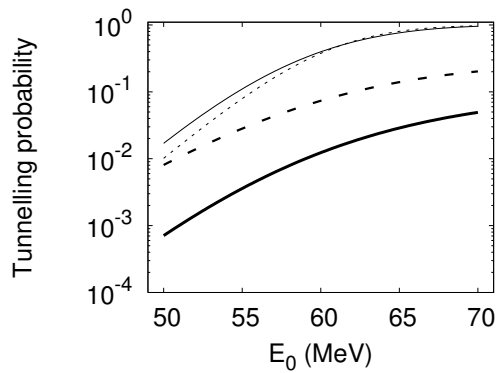}
\caption{Tunnelling probability through the Eckart potential barrier as a function of the initial total energy: analytic probabilities (dotted line), probabilities of the wave-packet with a definite energy spread (thin solid line), and the latter within the complex potential approach (thick solid line) and the Lindblad dissipative dynamics (dashed line). The environment reduces the tunnelling probability, playing decoherence a crucial role (comparing the thick solid line with the dashed line, and these with the thin solid line).} 
\label{Figure3}
\end{center}
\end{figure}

An interesting question arises: Does decoherence affect observables such as the quantum tunnelling probability? Yes, it does, as shown in Fig. \ref{Figure3}. Here, the tunnelling probability through the Eckart potential barrier is presented as a function of the initial total energy $E_0$. This probability is numerically calculated (after a long propagation time) as the trace of the density-matrix over negative values of the $x$ coordinate, \emph{i.e.}, Tr$(\hat{P} \hat{\rho})$ where the projector $\hat{P} = 
\hat{1} - \Theta (x)$, being $\Theta (x)$ the Heaviside step function. The dotted line represents the analytic transmission coefficient \cite{Eckart}, whilst the thin solid line shows the tunnelling probability for the wave-packet with a definite energy spread. This results in the small difference between the two lines. With respect to these, the tunneling probability is reduced in both the complex potential approach (thick solid line) and the Lindblad dynamics (dashed line). The difference between these two lines demonstrates that decoherence substantially changes the quantum tunnelling probability, \emph{i.e.}, up to a factor ten at the lowest incident energies. The energy dependence of this probability is clearly affected by decoherence (comparing the thick solid and dashed lines with the thin solid line). 
These results provide quantitative support to the idea \cite{Nanda1,Alexis1,Hinde1} that explicit inclusion of quantum decoherence may well be necessary to consistently describe (within a truncated model space) low-energy nuclear collision dynamics. 

Quantum decoherence is a key aspect of irreversibility, which is mostly overlooked in quantal models of low-energy nuclear phenomena. It should be treated simultaneously with energy dissipation and coherent quantum superpositions, the latter being the basis of the coherent coupled-channels approach to nuclear reaction dynamics around the Coulomb barrier. Coherent quantum superpositions manifest themselves through the measurement of the experimental fusion barrier distributions, while energy dissipation is revealed in deep-inelastic scattering that also occurs at near-barrier energies \cite{Hinde1}. While the coherent coupled channels calculations \cite{Balantekin1} are able to explain several collision observables, major problems are unresolved. Foremost is the inability to describe the elastic and quasi-elastic scattering and fusion processes simultaneously \cite{Newton1} and the related, more recent failure to describe consistently below-barrier quantum tunnelling and above-barrier fusion yields \cite{Nanda1}. New, precise measurements have inevitably led to phenomenological adjustments \cite{Esbensen1,Hagino1} (sometimes contradictory) to stationary-state coupled channels models to fit the experimental data, but without a physically consistent foundation. The present work suggests that quantum decoherence should be taken into account seriously in modelling fusion and nuclear scattering. This may also have significant implications for open systems-related approaches to phenomena in near-threshold exotic nuclei, such as continuum shell models \cite{Volya2} and approaches to understanding the quenching of spectroscopic factors 
\cite{Barbieri1,Timofeyuk1}.                       

In summary, the non-equivalence between the complex potential approach and the Lindblad dissipative dynamics for describing the time evolution of a schematic, open quantum system has been demonstrated. Time-dependent density-matrix propagation shows that quantum decoherence makes the difference, which is not treated in the complex potential model. It is inadequate for a consistent description of open quantum system dynamics. Decoherence substantially affects the quantum tunnelling probability, and should be included in a complete description of low-energy nuclear scattering.   

The author thanks Ron Johnson for stimulating discussions. This work was initiated by questions he raised. 
The author is also thankful to Gerard Milburn for illuminating conversations on the role of the imaginary potential, at the early stage of his research on decoherence in nuclear collision dynamics. 
Discussions with David Hinde, Nanda Dasgupta, Arnau Rios, Jim Al-Khalili and Jeff Tostevin are also acknowledged. The work was supported by the UK Science and Technology Facilities Council (STFC) Grant No. ST/F012012/1.

\end{document}